\documentclass[twocolumn,prd,floatfix,amsmath,nofootinbib,amssymb,floatfix]{revtex4}
\usepackage{graphicx,color,dcolumn,booktabs,bm}
\usepackage{longtable,lscape}
\usepackage{txfonts}
\usepackage{overpic}
\usepackage{amssymb}
\usepackage{indentfirst}
\usepackage{feynmf}
\usepackage{slashed}
\usepackage{cases}
\usepackage{multirow}
\usepackage[colorlinks,
            citecolor=blue,
            anchorcolor=red,
            menucolor=red,
            linkcolor=red,
            filecolor=red,
            runcolor=red,
            urlcolor=blue,
            frenchlinks=red]{hyperref}
\usepackage{epstopdf}
\usepackage{bm}
\usepackage{makecell}
\usepackage{threeparttable}
\begin{document}
\title{Possibility of charmoniumlike state $X(3915)$ as $\chi_{c0}(2P)$ state}
\author{Ming-Xiao Duan$^{1,2}$}\email{duanmx16@lzu.edu.cn}
\author{Si-Qiang Luo$^{1,2}$}\email{luosq15@lzu.edu.cn}
\author{Xiang Liu$^{1,2}$\footnote{Corresponding author}}\email{xiangliu@lzu.edu.cn}
\author{Takayuki Matsuki$^{3,4}$}\email{matsuki@tokyo-kasei.ac.jp}
\affiliation{
$^1$School of Physical Science and Technology, Lanzhou University, Lanzhou 730000, China\\
$^2$Research Center for Hadron and CSR Physics, Lanzhou University and Institute of Modern Physics of CAS, Lanzhou 730000, China\\
$^3$Tokyo Kasei University, 1-18-1 Kaga, Itabashi, Tokyo 173-8602, Japan\\
$^4$Theoretical Research Division, Nishina Center, RIKEN, Wako, Saitama 351-0198, Japan}

\begin{abstract}
In this work, we seriously discuss whether $X(3915)$ can be treated as a $\chi_{c0}(2P)$ state.
Based on an unquenched quark model, we give the mass spectrum of the $\chi_{cJ}(2P)$ states, where there are no free input parameters in our calculation. Our result shows that the mass gap between $\chi_{c0}(2P)$ and $\chi_{c2}(2P)$ can reach 13 MeV, which can reproduce the mass difference between $Z(3930)$ and $X(3915)$. Additionally, the calculated masses of $\chi_{c0}(2P)$ and $\chi_{c2}(2P)$ are consistent with experimental values of $X(3915)$ and $Z(3930)$, respectively. Besides, giving the mass spectrum analysis to support $X(3915)$ as
$\chi_{c0}(2P)$, we also calculate the width of  $\chi_{c0}(2P)$ with the same framework, which is also consistent with the experimental data of $X(3915)$. Thus, the possibility of charmoniumlike state $X(3915)$ as $\chi_{c0}(2P)$ state is further enforced.
\end{abstract}

\maketitle

\section{Introduction}\label{introduction}
As an important group of the whole hadron spectrum, the charmonium family plays a very important role to provide the hint for quantitatively understanding how quarks form different types of hadrons, which has a close relation to non-perturbative behavior of strong interactions. In 1974, the first charmonium state $J/\psi$ was found \cite{Aubert:1974js, Augustin:1974xw}. Then, in the subsequent eight years from 1974 to 1982, most of charmonia listed in the present Particle Data Group (PDG) were observed, which becomes the main body of the charmonium family. Here, the typical states include $J/\psi$ \cite{Aubert:1974js, Augustin:1974xw}, $\psi(3686)$ \cite{Abrams:1974yy}, $\psi(4040)$ \cite{Goldhaber:1977qn}, $\psi(4415)$ \cite{Siegrist:1976br}, $\psi(3770)$ \cite{Rapidis:1977cv}, $\psi(4160)$ \cite{Brandelik:1978ei}, $\chi_{c0}(1P)$ \cite{Biddick:1977sv}, $\chi_{c1}(1P)$ \cite{Tanenbaum:1975ef}, $\chi_{c2}(1P)$ \cite{Whitaker:1976hb}, $\eta_c(1S)$ \cite{Partridge:1980vk}, and $\eta_c(2S)$ \cite{Edwards:1981mq}. With these observations, the Cornel model was proposed by Eichten $et$ $al$. \cite{Eichten:1974af} in 1975, from which different versions of a potential model \cite{Krasemann:1979ir, Stanley:1980zm, Godfrey:1985xj, Radford:2007vd, Badalian:1999fe, Barnes:2005pb} applied to depict the interaction between quarks were developed by different groups.

\begin{center}
\begin{figure}[htbp]
\includegraphics[width=8.6cm,keepaspectratio]{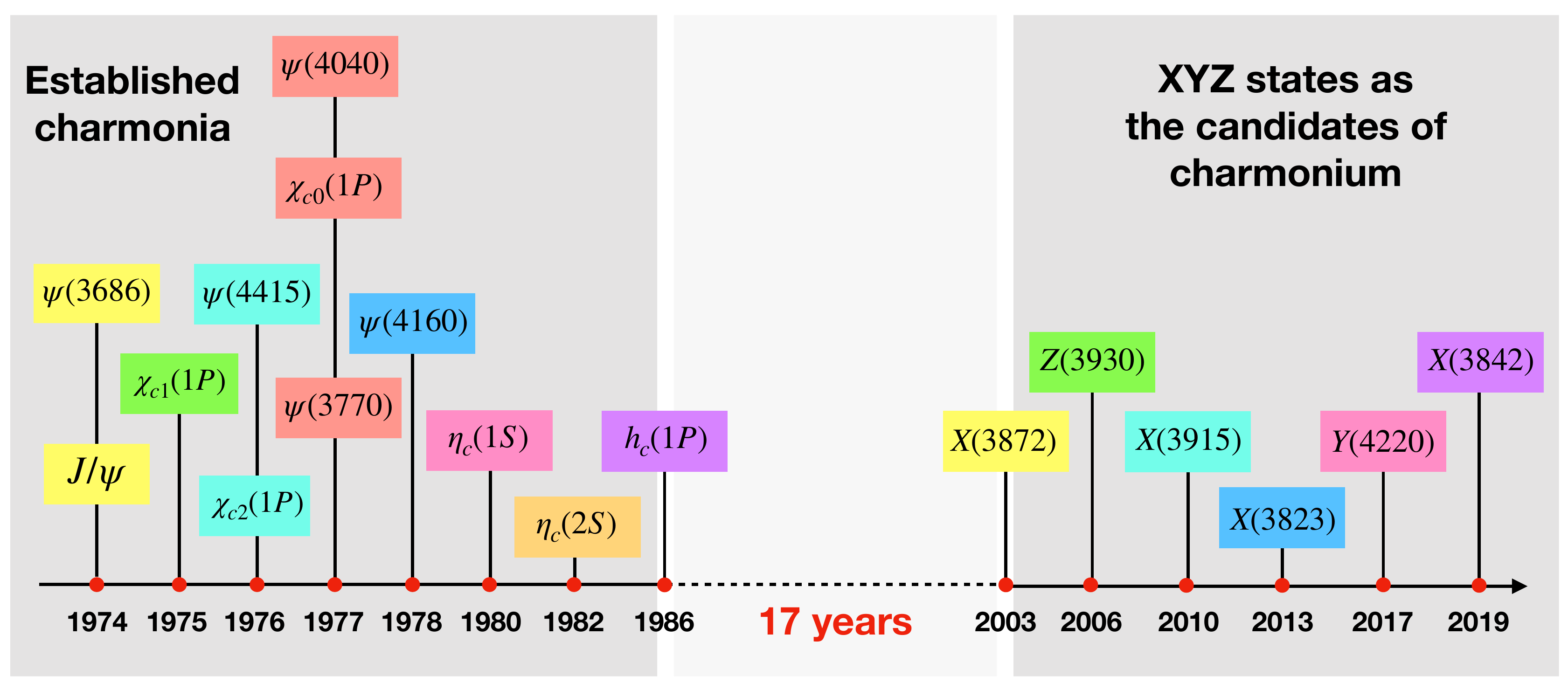}
\caption{The established charmonia and some $XYZ$ states as possible candidates for charmonium.}
\label{f1}
\end{figure}
\end{center}

However, the present observed charmonium spectrum is not complete in the sense that higher states in the charmonium family are still absent, where the higher states refer to the charmonia with higher radial and orbital quantum numbers. These missing higher states include three $1D$ states accompanied by $\psi(3770)$ and $2P$ states in the charmonium family. In fact, there is a big window {without discovery} of more new charmonia from 1982 to 2003, except $h_c$ reported by the R704 Collaboration \cite{Baglin:1986yd} in 1986. In Fig. \ref{f1}, all the observed charmonia and possible candidates are shown for the present status of charmonium family.

This situation has been dramatically changed as a series of charmoniumlike $XYZ$ states have been observed in experiments. $X(3872)$, as the first $XYZ$ states reported by the Belle collaboration \cite{Choi:2003ue}, stimulated theorists' interests in exploring $D\bar{D}^*$ molecular pictures \cite{Swanson:2003tb, Wong:2003xk, AlFiky:2005jd}, which has continued to date and shed light on the nature of $X(3872)$. For $X(3872)$, the experimental mass and decay width are measured as $M_{X(3872)}=$3.871 GeV and $\Gamma^{exp}_{X(3872)}<1.2$ MeV. The mass and width are far lower than predictions of potential models. By introducing coupled-channel effects, the low mass puzzle of $X(3872)$ can be well understood \cite{Barnes:2003vb, Ortega:2009hj, Kalashnikova:2005ui}. Thus, $X(3872)$ can be explained as a $\chi_{c1}(2P)$ state containing a $D\bar{D}^*$ component. And, two candidates of $1D$ states were announced by the Belle and LHCb Collaborations \cite{Bhardwaj:2013rmw, Aaij:2019evc}, which are $X(3823)$ from the $X(3823) \to \chi_{c1}\gamma$ decay channel and $X(3842)$ from the $X(3842) \to D\bar{D}$ process. In addition, the Lanzhou group indicated that there exists a narrow $Y$ state around 4.2 GeV, which corresponds to $\psi(4S)$ \cite{He:2014xna}. Later, BESIII indeed observed this narrow structure in the $e^+e^- \to \pi^+\pi^- h_c$ and $e^+e^- \to \omega\chi_{cJ}$ processes \cite{Chang-Zheng:2014haa, Ablikim:2014qwy}. Recently, they again published one paper to illustrate how to construct higher vector states of the $J/\psi$ family with updated data of charmoniumlike $Y$ states \cite{Wang:2019mhs}. From these examples, some of the charmoniumlike $XYZ$ states may be good candidates of missing charmonia. Thus, the above facts tell us a lesson, i.e., before introducing exotic hadronic state assignments to $XYZ$, we should carefully check whether there exists a possibility to group it into the charmonium family. Up to date, such a study has become an interesting research issue \cite{Barnes:2003vb, Kalashnikova:2005ui, Liu:2009fe, Chen:2012wy}.

In 2009, focusing on $2P$ states, the Lanzhou group carried out the study of the mass spectrum and strong decay behaviors of $2P$ charmonia by combining  the experimental data of $X(3872)$, $Z(3930)$, and $X(3915)$. Here, $Z(3930)$ and $X(3915)$ are from $\gamma\gamma\to D\bar{D}$ \cite{Uehara:2005qd} and $\gamma\gamma\to J/\psi\omega$ processes \cite{Uehara:2009tx}, respectively. Linking these $XZ$ states to charmonia, they indicated that $Z(3930)$ is the $\chi_{c2}(2P)$ state and decoded $X(3915)$ as the $\chi_{c0}(2P)$ state with definite $J^{PC}=0^{++}$ quantum number \cite{Liu:2009fe}. Later, the BaBar Collaboration confirmed this quantum number of $X(3915)$ \cite{Lees:2012xs}. Thus, $X(3915)$ as the $\chi_{c0}(2P)$ state was listed into the 2013 version of PDG \cite{Beringer:1900zz}.

After three years, this situation was changed by the paper  \cite{Guo:2012tv} with the title ``Where is the $\chi_{c0}(2P)$?". In this work, three questions were raised if treating $X(3915)$ as $\chi_{c0}(2P)$: 1) why $X(3915)\to J/\psi\omega$ has large width, 2) why the main decay mode ``$X(3915)\to D\bar{D}$"  was not reported in experiment, and 3) why the mass gap between $X(3915)$ and $Z(3930)$ is far smaller than that between $\chi_{b0}(2P)$ and $\chi_{b2}(2P)$. Then, two groups joined the discussion of whether $X(3915)$ can be the $\chi_{c0}(2P)$ state \cite{Olsen:2014maa, Olsen:2019lcx, Zhou:2015uva}. As a consequence, labeling $X(3915)$ as $\chi_{c0}(2P)$ was removed in the 2016 version of PDG \cite{Patrignani:2016xqp}.

Guo $et$ $al$. claimed that the $\chi_{c0}(2P)$ state should have mass around 3837.6$\pm$11.5 MeV and width about 221$\pm$19 MeV by their analysis to the $D\bar{D}$ invariant mass spectrum of the $\gamma\gamma\to D\bar{D}$ process \cite{Guo:2012tv}. In 2017, the Belle Collaboration made an analysis with $e^+e^-\to J/\psi D\bar{D}$ process, and found a broad structure named as $X(3860)$ \cite{Chilikin:2017evr}. Here, its mass and width are $M=3862$ MeV and $\Gamma=201$ MeV, respectively. Belle indicated that $X(3860)$ favors the $J^{PC}=0^{++}$ assignment. Therefore, Belle assigned the observed $X(3860)$ as $\chi_{c0}(2P)$.
In Ref. \cite{Ortega:2017qmg}, the authors studied charmoniumlike structures around 3.9 GeV in the framework of a constituent quark model.
Here, their result favors the hypothesis that $X(3915)$ and $Z(3930)$ resonances arise as different decay mechanisms of the same $J^{PC}=2^{++}$ state, and explained $X(3860)$ to be a $\chi_{c0}(2P)$ state \cite{Ortega:2017qmg}.

It is obvious that the situation of establishing the $\chi_{c0}(2P)$ candidate gets into a mess, which should be urgently clarified as soon as possible.

In the past years, we have been paying close attention to this problem. In Ref. \cite{Chen:2012wy}, the Lanzhou group proposed a solution to the second problem mentioned above. The structure corresponding to $Z(3930)$ observed in the $D\bar{D}$ decay channel may contain two $P-$wave higher charmonia $\chi_{c0}(2P)$ and $\chi_{c2}(2P)$, which can be supported by the analysis of the $D\bar{D}$ invariant mass spectrum and $\cos\theta^*$ distribution of $\gamma\gamma\to D\bar{D}$ \cite{Uehara:2005qd}. This means that the second problem raised in Ref. \cite{Guo:2012tv} can be solved. We suggest Belle II to reanalyze the $\gamma\gamma\to D\bar{D}$ process with more precise data.

We still believe that $X(3915)$ observed in $\gamma\gamma\to J/\psi\omega$ is a good candidate of $\chi_{c0}(2P)$. Thus, we must face the third problem raised in Ref.~\cite{Guo:2012tv} just mentioned above. In a quenched potential model, the mass splitting between $\chi_{c0}(2P)$ and $\chi_{c2}(2P)$ is far larger than that between $X(3915)$ and $Z(3930)$. According to the quenched quark model estimate, this relation $\mid m_{\chi_{c2}(2P)}-m_{\chi_{c0}(2P)}\mid$ $>$ $\mid m_{\chi_{b2}(2P)}-m_{\chi_{b0}(2P)}\mid$ can be naively obtained as claimed in Ref. \cite{Guo:2012tv}. In fact, we should be careful with this point. $X(3872)$ is a typical example, where there exists the low mass puzzle, {{i.e., the mass of $X(3872)$ is around 100 MeV lower than the value from the quenched quark model calculation \cite{Godfrey:1985xj}.}} This puzzle can be solved by a coupled-channel effect by calculating mass with an unquenched quark model \cite{Kalashnikova:2005ui}. In fact, for other $2P$ states which are above the threshold of open-charm decay channels, the coupled-channel effect should be seriously considered, which will be the task in this work. We will illustrate why the mass gap of $\chi_{c0}(2P)$ and $\chi_{c2}(2P)$ is far smaller than that of $\chi_{b0}(2P)$ and $\chi_{b2}(2P)$ by an unquenched quark model calculation. In the following sections, we will give a detailed illustration.

Finally, when treating $X(3915)$ as $\chi_{c0}(2P)$, we need to answer the remaining problem whether or not $\chi_{c0}(2P)$ has wide width, which is a crucial point we have to face. In this work, we will explicitly present that $\chi_{c0}(2P)$ should be a narrow state which is due to the node effect. Thus, two $\chi_{c0}(2P)$ candidates like $X(3840)$ in Ref. \cite{Guo:2012tv} and $X(3860)$ reported by the Belle Collaboration \cite{Chilikin:2017evr} should be excluded.

This paper is organized as follows. After the Introduction, we will introduce the mass problem of a quenched quark model. Next, we will give a coupled-channel picture for the discussed $\chi_{c0}(2P)$ state in Sec.~\ref{sec2}. In Sec.~\ref{sec3}, the numerical result will be presented. Especially, we give an analysis why we can get consistent results with experimental data of $X(3915)$.
At last, this paper ends with a summary in Sec.~\ref{sec5}.

\section{Mass problem of $2P$ charmonium states from quenched quark model}\label{sec2}

With the observation of a series of charmonia, the Cornell model for quantitatively depicting the strong interactions between quarks was proposed by Eichten {\it et al.}~\cite{Eichten:1974af}. Since then, different versions of a potential model were developed by different groups. Among them, the Godfrey-Isgur (GI) model \cite{Godfrey:1985xj} was extensively applied to study the hadron spectrum. In this work, we firstly illustrate the mass problem of a quenched quark model by presenting the spectrum of $2P$ charmonium states, where the GI model was adopted.\footnote{Here, we need to comment on the calculated result of the mass of $^{3}P_0$ $c\bar{c}$ state by the nonrelativistic quark model. In Ref.~\cite{Barnes:2005pb}, the authors adopted the nonrelativistic quark model to give the mass spectrum of the charmonium family. We may reproduce most of their results by applying a perturbation method, where $H_0$ and $H^\prime$ are treated as a solvable part and a perturbation term, respectively. However, for $1^3P_0$ and $2^3P_0$ states, the calculated masses are not stable and convergent when including higher order perturbation contributions. For example, if adopting the potential suggested in Ref.~\cite{Barnes:2005pb}, mass of $1^3P_0$ is 3.525, 3.425, 3.351, and 3.266 GeV and mass of $2^3P_0$ is 3.943, 3.854, 3.781 and 3.701 GeV when zeroth-order, first-order, second-order, and third-order perturbation contributions are considered step by step in calculation. If adopting the potential given in Ref.~\cite{Badalian:1999fe}, there exists the same problem for the calculation of the mass of $1^3P_0$ and $2^3P_0$ states. This problem is due to the singularity of $1/r^3$-like terms in the potential near $r=0$. However, in the GI model, this singularity is smeared. Thus, such a problem does not exist.}


The GI model is a semirelativistic potential model with a Hamiltonian
\begin{equation}\label{Hamiltonian}
\begin{split}
H=\sqrt{\textbf{p}^2+m_1^2}+\sqrt{\textbf{p}^2+m_2^2}+\tilde{V}(\textbf{p,r}),
\end{split}
\end{equation}
where $m_1$ and $m_2$ are masses of quark and antiquark. The potential $\tilde{V}(\textbf{p,r})$ is composed of a short-range $\gamma^\mu \otimes \gamma_\mu$ interaction of one-gluon exchange and a long-range $1\otimes 1$ linear color confining interaction. When taking the nonrelativistic limit, a familiar nonrelativistic potential can be obtained from $\tilde{V}(\textbf{p,r})$.
In the GI model, the relativistic corrections can be considered by smearing transformation and momentum-dependent factors.
Here, the smearing function should be introduced, i.e.,
\begin{eqnarray}
\rho_{ij}(\textbf{r}-\textbf{r}^\prime)=\frac{\sigma^3_{ij}}{\pi^{\frac{3}{2}}}e^{-\sigma_{ij}^2 (\textbf{r}-\textbf{r}^\prime)^2},
\end{eqnarray}
by which the confining potential $S(r)=br+c$ and one-gluon exchange potential $G(r)=-4\alpha_s(r)/(3r)$ can be smeared out by
\begin{equation}\label{smear}
\tilde{G}(r)(\tilde{S}(r))=\int d^3r^\prime \rho_{ij}(\textbf{r}-\textbf{r}^\prime) G(r^\prime)(S(r^\prime)).
\end{equation}
For a general relativistic form of the potential, it should be dependent on momenta of interacting quarks in the center-of-mass system. Thus, we should further modify this smeared potential $\tilde{V}(r)$ by
\begin{eqnarray}
\tilde{V}_i(r) \rightarrow\left(\frac{m_cm_{\bar{c}}}{E_c E_{\bar{c}}}\right)^{1 / 2+\epsilon_i}\tilde{V}_i(r)\left(\frac{m_cm_{\bar{c}}}{E_c E_{\bar{c}}}\right)^{1 / 2+\epsilon_i}
\end{eqnarray}
with $E_c=(p^2+m_c^2)^{1/2}$ and $E_{\bar{c}}=(p^2+m_{\bar{c}}^2)^{1/2}$, where a parameter $\epsilon_i$ corresponds to different types of interactions. The details of the GI model can be found in Ref. \cite{Godfrey:1985xj}.

\begin{table}[htbp]
\caption{The parameters involved in the GI model and their values by fitting the well-established charmonia.}
\label{tablepara}
\renewcommand\arraystretch{1.20}
\begin{tabular*}{80mm}{@{\extracolsep{\fill}}cccccc}
\toprule[1pt]
\toprule[1pt]
$m_q$  & 0.220 GeV& $b$ & 0.175                 & $\epsilon_{\rm cont}$&-0.103\\
$m_s$  & 0.419 GeV& $\alpha^{\rm critical}_s$ & 0.6& $\epsilon_{\rm tens}$&-0.114\\
$m_c$  & 1.628 GeV& $\Lambda$    & 200 MeV     & $\epsilon_{\rm so(v)}$&-0.279\\
$s$    & 0.821 GeV & $c$    & -0.245 GeV        & $\epsilon_{\rm so(s)}$&-0.3\\
       &          & $\sigma_0$    & 2.33 GeV    & \\
\bottomrule[1pt]
\bottomrule[1pt]
\end{tabular*}
\end{table}

\begin{center}
\begin{figure}[htbp]
\includegraphics[width=8.1cm,keepaspectratio]{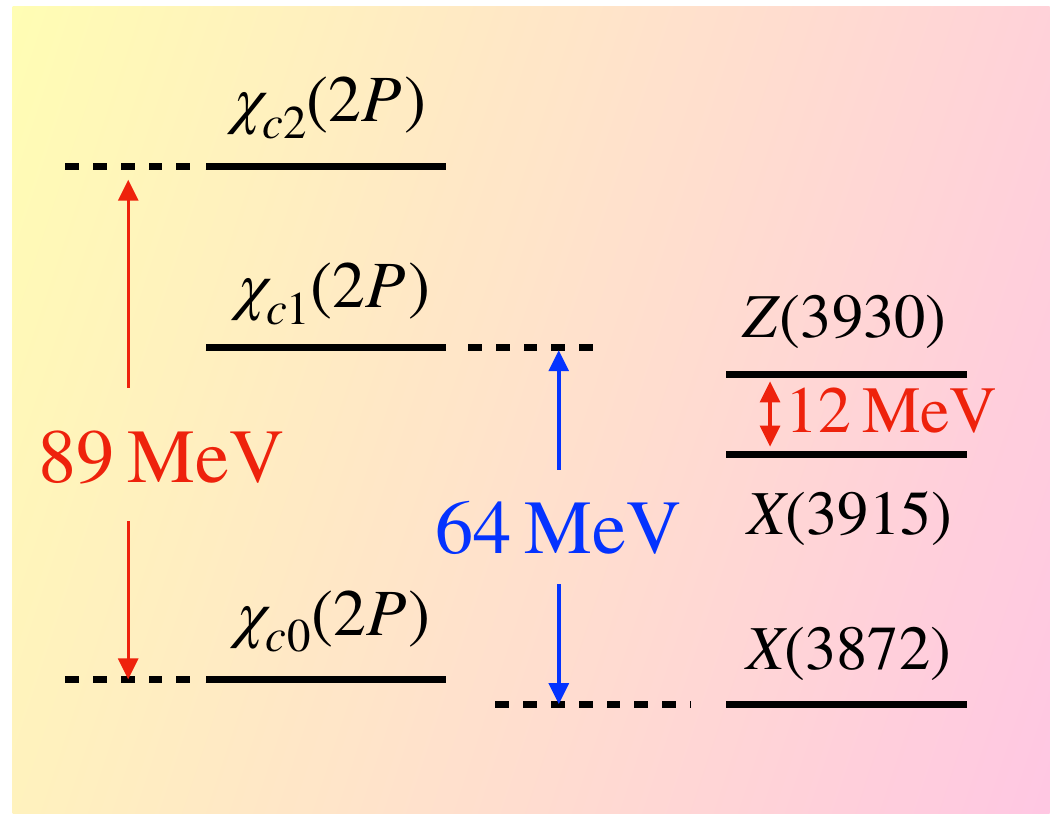}
\caption{The masses of spin triplet of $2P$ charmonia given by the GI model and the comparison with three charmoniumlike states $X(3872)$, $X(3915)$ and $Z(3930)$. Here, the $J^P$ quantum numbers of $X(3872)$ and $X(3915)$ were measured in experiment which are $1^{++}$ \cite{Aaij:2013zoa} and $0^{++}$ \cite{Lees:2012xs}, respectively. }
\label{tu2}
\end{figure}
\end{center}

In Table~\ref{tablepara}, we list the parameters of the GI model, which can be obtained by refitting the masses of the low-lying well-established charmonia ($\eta_c(1S)$, $J/\psi$, $\psi(3686)$, $\psi(3770)$, $h_c(1P)$, $\chi_{c0}(1P)$, $\chi_{c1}(1P)$, $\chi_{c2}(1P)$, $\psi(4040)$, and $\psi(4160)$)~\cite{Tanabashi:2018oca}. The obtained values are slightly different from those given in Ref.~\cite{Godfrey:1985xj}.
Here, the obtained masses (in units of GeV) of $1^1S_0$, $1^3S_1$, $2^1S_0$, $2^3S_1$, $3^3S_1$, $1^1P_1$, $1^3P_0$, $1^3P_1$, $1^3P_2$, $1^3D_1$, $1^3D_2$, $1^3D_3$, and $2^3D_1$ are $2.996$, $3.098$, $3.634$, $3.676$, $4.090$, $3.513$, $3.417$, $3.500$, $3.549$, $3.805$, $3.828$, $3.841$, and $4.172$, respectively. Just shown in above, these low-lying charmonia can be well reproduced.

With the same parameters as input, we may give the masses of $2P$ states and make a comparison with the observed $X(3872)$, $X(3915)$, and $Z(3930)$. There exists the 64 MeV difference between $2^3P_1$ charmonium and $X(3872)$, which is the famous low mass puzzle of $X(3872)$. In addition, the mass gap (89 MeV) between $2^3P_0$ and $2^3P_2$ $c\bar{c}$ states
is far larger than that between $X(3915)$ and $Z(3930)$, which is 12 MeV. In Fig.~\ref{tu2}, the difference of mass spectrum between the $2P$ states given by the GI model and the observed three charmoniumlike states is explicitly illustrated.

This is the mass problem of the $2P$ charmonium spectrum by the quenched quark model. Hence, we should develop an unquenched picture when facing such a mass problem since the allowed open-charm decay channels are open for these $2P$ states. This will be the crucial task dedicated in this paper.

\section{The mass spectrum of $2P$ charmonia by an unquenched picture}\label{sec3}

When checking the masses from a quenched quark model like the GI model, we notice that the discussed $2P$ $c\bar{c}$ states are above the $D\bar{D}$ and $D\bar{D}^*$ thresholds. For $\chi_{c1}(2P)$, $S$-wave and $D$-wave interactions occur for the $\chi_{c1}(2P)$ coupling with the $D\bar{D}^*$. For $\chi_{c0}(2P)$, it can couple with $D\bar{D}$ via an $S$-wave interaction while $\chi_{c2}(2P)$ may interact with the $D\bar{D}$ and $D\bar{D}^*$ via a $D$-wave coupling. Thus, in this section we exam the coupled-channel effect from the $D\bar{D}$ and $D\bar{D}^*$ channels to the mass spectrum of $2P$ charmonia. In the following subsection, we first introduce some historical results of $\chi_{c0}(2P)$ presented in some published literatures. After that, the unquenched model adopted in this paper will be introduced.

\subsection{The research status of mass of $\chi_{c0}(2P)$ and  $\chi_{c2}(2P)$}\label{puzzle}

In fact, there were some theoretical papers of the calculation of mass of $\chi_{c0}(2P)$ and $\chi_{c2}(2P)$ states
under the unquenched picture \cite{Kalashnikova:2005ui, Pennington:2007xr, Zhou:2013ada, Li:2009ad, Ono:1983rd} before the present work, which are summarized in Table \ref{massshift}.
\begin{table}[htbp]
\footnotesize
\caption{Mass of $\chi_{c0}(2P)$ and $\chi_{c2}(2P)$ states from different theoretical groups. Here, the bare and physical masses and the corresponding mass shift are collected. }
\label{massshift}
\begin{threeparttable}
\renewcommand\arraystretch{1.00}
\begin{tabular*}{86mm}{@{\extracolsep{\fill}}c|ccc|ccc}
\toprule[1pt]
\toprule[1pt]
       &            &$\chi_{c0}(2P)$       &        &           &$\chi_{c2}(2P)$            &\\
   Ref.&$m_{\rm bare}$&$m_{\rm phy}$&mass shift&$m_{\rm bare}$&$m_{\rm phy}$&mass shift  \\
\toprule[0.8pt]
\cite{Kalashnikova:2005ui}&4108&3918\tnote{1}&-190&4230&3990&-240\\
\cite{Pennington:2007xr}&3852&3782\tnote{2}&-70&3972&3917&-55\\
\cite{Zhou:2013ada}&3916&3814\tnote{2}&-102&3979&3942&-37\\
\cite{Li:2009ad}&3948&3915\tnote{1}&-33&4085&3966&-119\\
\cite{Ono:1983rd}&3990&3893\tnote{1}&-97&4104&3957&-147\\
\bottomrule[1pt]
\bottomrule[1pt]
\end{tabular*}
\begin{tablenotes}
\footnotesize
\item[1]The $D\bar{D}$, $D\bar{D}^*$, $D^*\bar{D}^*$, $D_s\bar{D}_s$, $D_s\bar{D}_s^*$, $D_s^*\bar{D}_s^*$ channels are contained in their calculations. The bare mass is gotten from a mass spectrum, where the contributions from the above channels are subtracted.
\item[2]Only the open channels are considered in these papers. The bare masses are gotten from the potential model fitted with experimental mass directly.
\end{tablenotes}
\end{threeparttable}
\end{table}

The results in Table \ref{massshift} show that the effect from open-charm channel contributions to the mass of $\chi_{c0}(2P)$ and $\chi_{c2}(2P)$ are obvious. However, if checking the details of the obtained results, inconsistency\footnote{We also notice the result in Ref.~\cite{Danilkin:2010cc} which is not listed in Table \ref{massshift}, where the $D\bar{D}$ channel only gives a 2 MeV contribution to the mass shift of $\chi_{c0}(2P)$.} still exists in the results. Especially, the small mass gap between $X(3915)$ and $Z(3930)$ in Fig. \ref{tu2} cannot be reproduced exactly. According to the general physical picture, we may conclude that the S-wave coupled-channel contribution to the mass shift should be larger than the D-wave coupled-channel, 
which in fact was not reflected by some concrete results in Refs.~\cite{Kalashnikova:2005ui, Li:2009ad, Ono:1983rd}. To some extent, the authors in Refs.~\cite{Li:2009ad, Ono:1983rd} did not realize this problem. Thus, the messy situation of mass study of $\chi_{c0}(2P)$ and $\chi_{c2}(2P)$ should be clarified by a more in-depth research, which is the main task of the present work.

\subsection{The adopted unquenched model}\label{model}

\begin{center}
\begin{figure}[htbp]
\includegraphics[width=6.3cm,keepaspectratio]{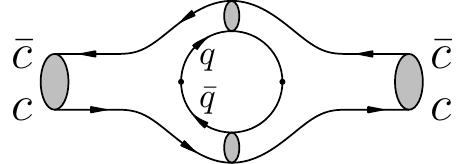}
\caption{The self-energy hadronic loop correction to $2P$ charmonium states. Here, $q=u,d,s$ and the intermediate loops are composed of charmed or charmed-strange mesons.}
\label{loop}
\end{figure}
\end{center}

The description of self-energy hadronic loop corrections to $2P$ charmonium states is illustrated in Fig. \ref{loop}. Here, a bare state can be dressed by these coupled hadronic channels composed of charmed mesons, which corresponds to a physical state.

For giving a quantitative calculation for it, we need to construct the coupled-channel equation
\begin{equation}\label{inversepropagator}
\textbf{P}^{-1}(s)\equiv m_{\rm bare}^2-s+\Pi(s)=0,
\end{equation}
where the $m_{\rm bare}$ is the mass of a bare state which can be calculated by a quenched quark model like the GI model as described in Sec.~\ref{sec2}.
$s$ is a pole found in a complex energy plane. The $\Pi(s)$ is the summation of $\Pi_n(s)$, and the subscript $n$ in $\Pi_n(s)$ denotes the $n$-th hadronic channel coupled with this bare $c\bar{c}$ state. The $s$ fulfilling the $\textbf{P}^{-1}=0$ is the coupled-channel result. The $s$ is defined as $s=(m_{\rm phy}-{\rm i}{\Gamma}/{2})^2$, where $m_{\rm phy}$ and $\Gamma$ are the mass and width of a physical state which may correspond to
experimental resonance parameters of the concrete observed state.

For a discussed heavy quarkonium, the narrow width approximation $s\approx m_{\rm phy}^2-{\rm i}m_{\rm phy}\Gamma$ can be employed in Eq.~(\ref{inversepropagator}). Then, the real and imaginary parts of Eq.~(\ref{inversepropagator}) can be separated, i.e.,
\begin{equation}\label{Narrow}
\begin{split}
m_{\rm phy}^2=&m_{\rm bare}^2+{\rm Re}\Pi(m_{\rm phy}^2),\\
\Gamma=&-\frac{{\rm Im}\Pi(m_{\rm phy}^2)}{m_{\rm phy}},
\end{split}
\end{equation}
from which $m_{\rm phy}$ and $\Gamma$ are directly calculated. By solving the first equation in Eq. (\ref{Narrow}), $m_{\rm phy}$ can be obtained, which can be subsequently applied to get the width $\Gamma$ by the second equation in Eq.~({\ref{Narrow}}).

Using the optical theorem, the imaginary part ${\rm Im}\Pi_n(m_{\rm phy}^2)$ in Eq. (\ref{Narrow}) can be calculated by cutting the hadronic loop shown in Fig.~\ref{loop}. The interaction between a bare state and a hadronic channel is described by an amplitude $M^{LS}(P)$, which has a close relation with the imaginary part ${\rm Im}\Pi_n(m_{\rm phy}^2)$ \cite{Barnes:2007xu}, i.e.,
\begin{equation}\label{ImPi}
{\rm Im}\Pi_n(m_{\rm phy}^2)=-2\pi PE_BE_C |M^{LS}(P)|^2,
\end{equation}
where $B$ and $C$ are two intermediate mesons which are the components of a constructing hadronic loop. $P$ represents 
the momentum of a $B$ meson. Using the K\"allen function $\lambda(x,y,z)=x^2+y^2+z^2-2xy-2xz-2yz$, the momentum $P$ can be expressed as $P=\lambda^{1/2}(m_{\rm phy}^2,m_B^2,m_C^2)/(2m_{\rm phy})$. Then, $M^{LS}(P)$ can be transferred into $M^{LS}(m_{\rm phy})$ which will be abbreviated as $M^{LS}$ for convenience. $E_B$ and $E_C$ are energies of $B$ and $C$ mesons, which can be represented as $E_{B/C}=\sqrt{P^2+m^2_{B/C}}$. The amplitude $M^{LS}$ can be given by the quark pair creation (QPC) model \cite{Micu:1968mk, LeYaouanc:1972vsx, Blundell:1996as, Ackleh:1996yt}, which will be explicitly introduced later.

Next, the corresponding real part ${\rm Re}\Pi_n(m^2_{\rm phy})$ can be related to the imaginary part ${\rm Im}\Pi_n(m_{\rm phy}^2)$ by the dispersion relation,
\begin{equation}\label{RePi}
{\rm Re}\Pi_n(m_{\rm phy}^2)=\frac{1}{\pi}\mathcal{P}\int^{\infty} _{S_{{\rm th},n}} {\rm d}z\frac{{\rm Im}\Pi_n(z)}{z-m_{\rm phy}^2}.
\end{equation}
Here. the $\mathcal{P}$ denotes of principal value integration, and $S_{{\rm th},n}$ is the threshold of the $n-$th channel.

Notice that because of the optical theorem, we could sum over the contributions from all possible intermediate hadronic loops, if Eq.~(\ref{RePi}) is used. However, this treatment is not realistic, which is a problem if directly applying Eq.~(\ref{RePi}) to calculate the coupled-channel correction to the bare mass. For solving this problem, the once subtracted dispersion relation was proposed in Ref.~\cite{Pennington:2007xr} by Pennington {\it et al.}. In this work, we employ this once subtracted ${\rm Re}\Pi_n(m_{\rm phy}^2)$
\begin{equation}\label{subPi}
{\rm Re}\Pi_n(m_{\rm phy}^2)=\frac{m_{\rm phy}^2-m^2_0}{\pi}\mathcal{P}\int^{\infty} _{S_{{\rm th},n}} {\rm d}z\frac{{\rm Im}\Pi_n(z)}{(z-m_{\rm phy}^2)(z-m^2_0)},
\end{equation}
where the subtraction point $m_0$ may correspond to a ground state, which is usually much lower than the threshold of
the first OZI-allowed coupled channel. For a discussed charmonium system, we may choose the mass of $J/\psi$ particle ($m_{J/\psi}=$3.097 GeV) as $m_0$. With this subtraction method given in Eq.~(\ref{subPi}), only the hadronic channels whose thresholds are lower than the mass of a discussed bare state are taken into consideration, by which the coupled-channel corrections become calculable.

In the following, we should briefly introduce how to employ the QPC model to get the partial wave amplitude $M^{LS}$ appearing in Eq.~(\ref{ImPi}). In the QPC model, a transition operator $\hat{T}$ is defined as \cite{Blundell:1996as}
\begin{equation}\label{Tmatrix}
\begin{split}
\hat{T}=&-3 \gamma \sum_{m}\langle 1,m;1,-m |0,0\rangle \int {\rm d}^{3} {\bf p}_{3} {\rm d}^{3} {\bf p}_{4}\;\delta^{3}({\bf p}_{3}+{\bf p}_{4})\\
  &\times\mathcal{Y}_{1}^{m}(\frac{{\bf p}_{3}-{\bf p}_{4}}{2})\chi_{1-m}^{34} \phi_0^{34} \omega_0^{34} b_{3}^{\dagger}({\bf p}_{3})d_{4}^{\dagger}({\bf p}_{4}),
\end{split}
\end{equation}
where ${\bf p}_3$ and ${\bf p}_4$ are momenta of the quark and antiquark, respectively, which are created from the vacuum. $b_{3}^{\dagger}$ and $d_{4}^{\dagger}$ represent the quark and antiquark creation operators. $\chi^{34}$, $\phi_0^{34}$, $\omega_0^{34}$, and $\mathcal{Y}_{1}^{m}$ are spin, flavor, color, and orbital wave functions of the created quark pair, respectively. The $\gamma$ depicts the strength of a quark-antiquark pair created from the vacuum, which is fixed by fitting the experimental data. Finally, the $M^{LS}$ could be expressed as
\begin{eqnarray}
\label{MLS}
&&M^{LS}\nonumber\\
&&=3\gamma \frac{\sqrt{4\pi(2L+1)}}{2J_A+1}\sum\limits_{M_{J_B}M_{J_C}}\langle L0S(M_{J_B}+M_{J_C})|J_A(M_{J_B}+M_{J_C})\rangle\nonumber\\&&\quad\times
\langle J_BM_{J_B}J_CM_{J_C}|S(M_{J_B}+M_{J_C})\rangle\nonumber \\
&&\quad\times \langle L_{A} M_{L_{A}}S_{A} M_{S_{A}} | J_{A} (M_{J_{B}}+M_{J_C})\rangle
\nonumber\\&&\quad\times\sum\limits_{\mbox{\tiny{$\begin{array}{c}{M_{L_{A}},M_{S_{A}},M_{L_B},M_{S_B}}\\M_{L_C},M_{S_C},m\end{array}$}}}\langle L_{A} M_{L_{A}} S_{A} M_{S_{A}} | J_{A} (M_{J_{B}}+M_{J_C})\rangle\nonumber\\
&&\quad\times\langle L_{B} M_{L_{B}} S_{B} M_{S_{B}} | J_{B} M_{J_{B}}\rangle\langle L_{C} M_{L_{C}} S_{C} M_{S_{C}} | J_{C} M_{J_{C}}\rangle \nonumber\\&&\quad\times
\langle 1,m;1,-m |0,0\rangle\langle \chi_{S_BM_{S_B}}^{14}\chi_{S_CM_{S_C}}^{32}|\chi_{S_AM_{S_A}}^{12}\chi_{1-m}^{34}\rangle
\nonumber\\&&\quad\times
\langle\omega^{14}_{B} \omega^{32}_C| \omega^{12}_A \omega_0^{34} \rangle\left[\langle\phi_{B}^{14} \phi_{C}^{32} | \phi_{A}^{12} \phi_{0}^{34}\rangle I(P\hat{\bf z}, m_{1}, m_{2}, m_{3})\right.\nonumber\\&&\quad\left.+(-1)^{1+S_{A}+S_{B}+S_{C}}\langle\phi_{B}^{32} \phi_{C}^{14} | \phi_{A}^{12} \phi_{0}^{34}\rangle I(-P\hat{\bf z}, m_{2}, m_{1}, m_{3})\right].
\end{eqnarray}
Here, the integral $I(P\hat{\bf z},m_1,m_2,m_3)$ is the overlap of the finial and initial wave functions in momentum space
\begin{equation}\label{int}
\begin{split}
I(P\hat{\bf z}, m_{1}, m_{2}, m_{3})=& \int \mathrm{d}^{3} {\bf p}\; \psi_{n_{B} L_{B} M_{L_{B}}}^{*}\left({\bf p}-\frac{m_{1}}{m_{1}+m_{3}} P\hat{\bf z}\right)\\
&\times\psi_{n_{C} L_{C} M_{L_C}}^{*}\left({\bf p}-\frac{m_{2}}{m_{2}+m_{3}} P\hat{\bf z}\right)\\
&\times\mathcal{Y}_{1}^{m}({\bf p}-P\hat{\bf z})\psi_{n_{A} L_{A} M_{L_A}}({\bf p}),
\end{split}
\end{equation}
where $\psi_{nLM}({\bf p})$ is the spatial wave function of a meson state, which can be given by the GI model. It could be decomposed as
$\psi_{nLM}({\bf p})=R_{nL}(p)Y_{LM}(\hat{\bf p})$,
where the numerical result of $R_{nL}(p)$ for the involved mesons will be given in the next subsection and $Y_{LM}(\hat{\bf p})$ represents the angular part. 

With these preparations, we will present the numerical results in the next subsection.

\subsection{The numerical results}\label{dissect}

To present the numerical result, the key point is to quantitatively calculate a bare $c\bar{c}$ $2P$ state coupling with the corresponding open-charm channels.
As described in Sec. \ref{model}, the $\gamma$ value should be provided, and spatial wave functions of charmonia and charmed mesons involved in this work should be given.

\begin{figure}[htbp]
\centering
\includegraphics[width=8.6cm,keepaspectratio]{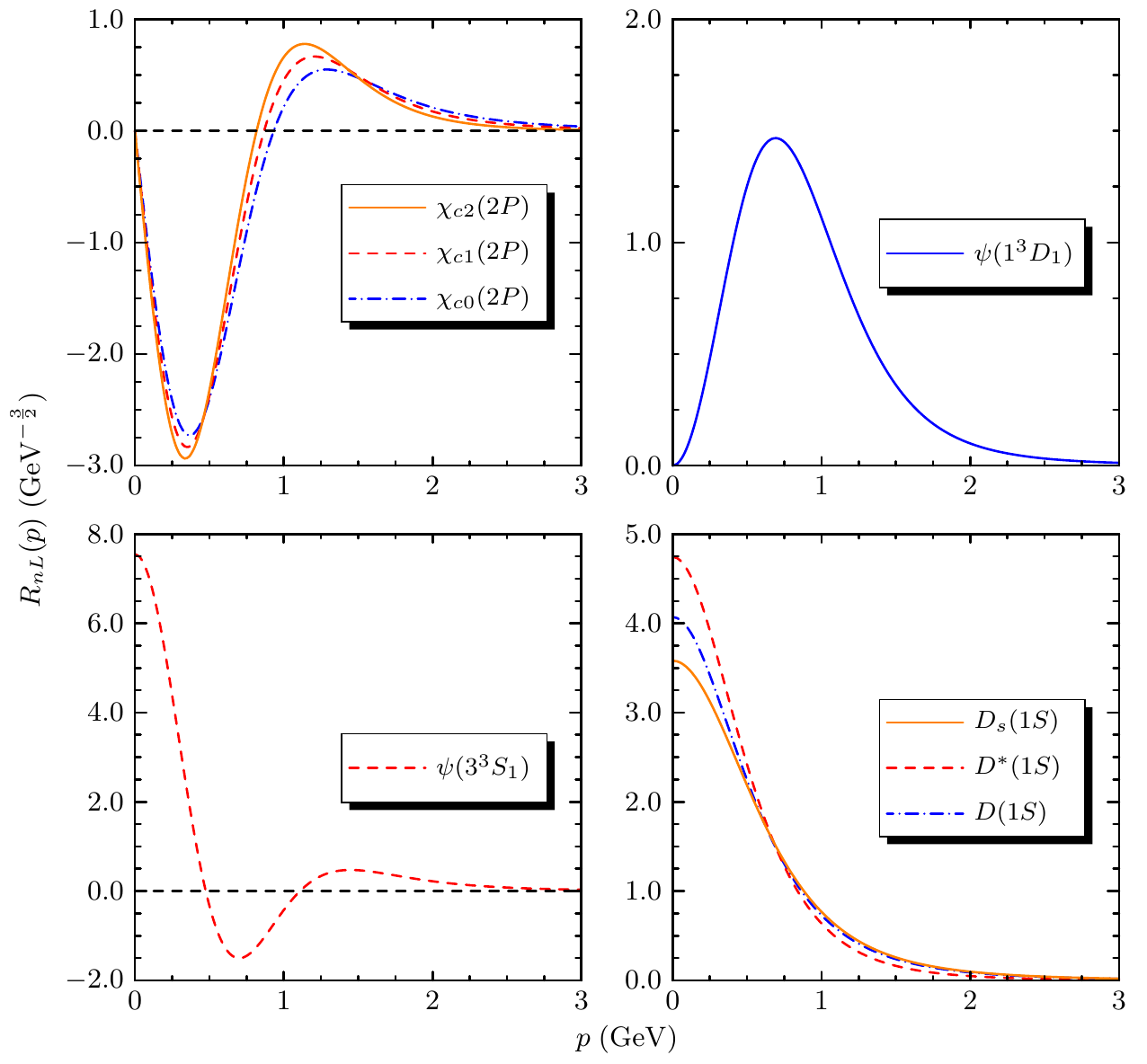}
\caption{The radial wave functions of the involved mesons from the GI model calculation in Sec.~\ref{sec2}. Here, the factor $(-{\rm i})^L$ is omitted, which does not affect the physical results in this work.}
\label{radial}
\end{figure}

As shown in Sec. \ref{sec2}, the numerical spacial wave functions of the mesons involved in this work can be obtained with the help of the GI model, where the numerical results of a radial part $R_{nL}(p)$ for the involved mesons are collected in Fig.~\ref{radial}.

Instead of directly applying the obtained numerical radial wave functions to concrete calculation, 
we adopt $R_{nL}(p)=\sum_{n=1}^{n_{max}}C_{n}\mathcal{R}_{nL}^{\rm SHO}(p)$, where
$\mathcal{R}_{nL}^{\rm SHO}$ is the simple harmonic oscillator (SHO) basis
with an expression
\begin{eqnarray}
&&\mathcal{R}_{nL}^{\rm SHO}({ p})\nonumber\\
&&=\frac{(-1)^{n-1}(-i)^L}{\beta^{\frac{3}{2}}}\sqrt{\frac{2(n-1)!}
{\Gamma(n+L+\frac{1}{2})}}\left(\frac{p}{\beta}\right)^Le^{-\frac{p^2}{2\beta^2}}L_{n-1}^{L+\frac{1}{2}}\left(\frac{p^2}{\beta^2}\right),\nonumber\\\label{wav0}
\end{eqnarray}
For different states, we choose $\beta=0.5$ and $n_{max}=20$, by which the numerical wave functions shown in Fig. \ref{radial} can be well reproduced. Here, the values of $C_n$ $(n=1-20)$ are collected in Tables~\ref{C1}-\ref{C2}.

\begin{table}[htbp]
\footnotesize
\caption{The values of $C_n$ $(n=1,2,\cdots,20)$ to reproduce the numerical radial wave functions of $\chi_{cJ}(2P)$ and $\psi(1^3D_1)$ in Fig. \ref{radial}.}
\label{C1}
\renewcommand\arraystretch{1.1}
\begin{tabular*}{86mm}{@{\extracolsep{\fill}}ccccc}
\toprule[1pt]
\toprule[1pt]
$C_n$ & $\chi_{c0}(2P)$ & $\chi_{c1}(2P)$ & $\chi_{c2}(2P)$ & $\psi(1^3D_1)$\\
\toprule[0.8pt]
$C_1$ &-0.4143005333&-0.2843674639&-0.1676617871&0.9774736067\\
$C_2$ &0.8404062724&0.9214858898&0.9698447346&0.1358246808\\
$C_3$ &0.1889966268&0.1226379196&0.0355260912&0.1368425228\\
$C_4$ &0.2206650135&0.1943672207&0.1608808564&0.0586720974\\
$C_5$ &0.1187442290&0.0814078379&0.0385803776&0.0443341708\\
$C_6$ &0.0972576561&0.0707988554&0.0421864384&0.0282505346\\
$C_7$ &0.0692867953&0.0453031791&0.0198853318&0.0212717487\\
$C_8$ &0.0553864904&0.0359848223&0.0159988286&0.0157696071\\
$C_9$ &0.0436391753&0.0269896878&0.0100372317&0.0123820607\\
$C_{10}$ &0.0357426112&0.0216974016&0.0075839604&0.0098380167\\
$C_{11}$ &0.0294901747&0.0174196570&0.0053734274&0.0080194428\\
$C_{12}$ &0.0247632590&0.0143745749&0.0040706414&0.0066226763\\
$C_{13}$ &0.0209684539&0.0119603710&0.0030550108&0.0055457695\\
$C_{14}$ &0.0179500110&0.0100906442&0.0023420753&0.0046949549\\
$C_{15}$ &0.0154172210&0.0085715809&0.0018238133&0.0039968502\\
$C_{16}$ &0.0134312438&0.0073601625&0.0013873140&0.0034541240\\
$C_{17}$ &0.0114735863&0.0062784400&0.0011435597&0.0029304188\\
$C_{18}$ &0.0104085833&0.0055533857&0.0007883739&0.0026436853\\
$C_{19}$ &0.0080198357&0.0044020432&0.0007780672&0.0020248026\\
$C_{20}$ &0.0091080535&0.0046668322&0.0003497855&0.0022979575\\
\bottomrule[1pt]
\bottomrule[1pt]
\end{tabular*}
\end{table}

\begin{table}[htbp]
\footnotesize
\caption{The values of $C_n$ $(n=1,2,\cdots,20)$ to reproduce the numerical radial wave functions of $\psi(3^3S_1)$ and charmed mesons in Fig. \ref{radial}.}
\label{C2}
\renewcommand\arraystretch{1.1}
\begin{tabular*}{86mm}{@{\extracolsep{\fill}}ccccc}
\toprule[1pt]
\toprule[1pt]
$C_n$ &  $\psi(3^3S_1)$ & $D$ & $D^*$ & $D_s$\\
\toprule[0.8pt]
$C_1$ &-0.0992718502&0.9572904583&0.9865559279&0.9443017126\\
$C_2$ &-0.3374923597&0.1825918937&0.0680481013&0.2307929570\\
$C_3$ &0.8955788540&0.1817331834&0.1360498850&0.1813594151\\
$C_4$ &0.0617570803&0.0801633067&0.0310250496&0.0967093768\\
$C_5$ &0.2255541433&0.0728160884&0.0421465440&0.0749847869\\
$C_6$ &0.0768962829&0.0430908329&0.0156328977&0.0510236218\\
$C_7$ &0.0825487703&0.0382417626&0.0181746661&0.0405972604\\
$C_8$ &0.0487249825&0.0260513695&0.0086611075&0.0307169610\\
$C_9$ &0.0412764414&0.0229624531&0.0093519699&0.0250604139\\
$C_{10}$ &0.0300564574&0.0169590578&0.0051437458&0.0200640064\\
$C_{11}$ &0.0245477609&0.0148880186&0.0053563851&0.0166755766\\
$C_{12}$ &0.0194363155&0.0116148335&0.0032086072&0.0138410163\\
$C_{13}$ &0.0160355493&0.0101167361&0.0032967040&0.0116231318\\
$C_{14}$ &0.0131933394&0.0082650453&0.0020670878&0.0099353357\\
$C_{15}$ &0.0110759571&0.0070565574&0.0021411336&0.0083202705\\
$C_{16}$ &0.0092823524&0.0060863301&0.0013498963&0.0073784008\\
$C_{17}$ &0.0079033729&0.0049219290&0.0014562009&0.0059660834\\
$C_{18}$ &0.0066896346&0.0046946936&0.0008680175&0.0057311122\\
$C_{19}$ &0.0055357701&0.0031600071&0.0010199307&0.0039539041\\
$C_{20}$ &0.0052012926&0.0040956058&0.0005608933&0.0050360564\\
\bottomrule[1pt]
\bottomrule[1pt]
\end{tabular*}
\end{table}

To determine the $\gamma$ value, we need to reproduce the widths of $\psi(3770)$ and $\psi(4040)$, which are treated as $\psi(1^3D_1)$ and $\psi(3^3S_1)$ charmonium states, respectively.
The allowed open-charm decay channels are the $D\bar{D}$ mode for $\psi(3770)$, and the $D\bar{D}$, $D\bar{D}^*$, $D^*\bar{D}^*$, and $D_sD_s$ modes for $\psi(4040)$, where the sum of these open-charm decays almost provides the width of these two charmonia.
The QPC model is employed to calculate the corresponding partial decay widths (the details of the QPC model can be found in Eqs. (\ref{Tmatrix})-(\ref{MLS}))\footnote{The expression of width is
\begin{equation}
\Gamma=2\pi\frac{PE_BE_C}{m_{\rm phy}}\sum_{LS}\left|M^{LS}(P)\right|^2,
\end{equation}
which is equivalent to $\Gamma$ in the second equation in Eq.~(\ref{Narrow}). Here, $M^{LS}$
is given by
Eq. (\ref{MLS})}. We find that taking $\gamma=0.4$, the experimental width of $\psi(3770)$ and $\psi(4040)$ ($\Gamma_{\psi(3770)}^{exp}=$27.2 MeV and $\Gamma_{\psi(4040)}^{exp}=$80 MeV \cite{Tanabashi:2018oca}) can be reproduced here. In this calculation, the obtained numerical wave functions shown in Fig. \ref{radial} and Tables \ref{C1}-\ref{C2} are input.
Additionally, we give the masses of the involved states $\psi(3770)$, $\psi(4040)$, $D$, $D^*$, and $D_s$ as $m_{\psi(3770)}=3.773$ GeV, $m_{\psi(4040)}=4.039$ GeV, $m_D=1.867$ GeV, $m_{D^*}=2.009$ GeV, and $m_{D_s}=1.968$ GeV, respectively.

\begin{figure}[htbp]
\centering
\includegraphics[width=8.6cm,keepaspectratio]{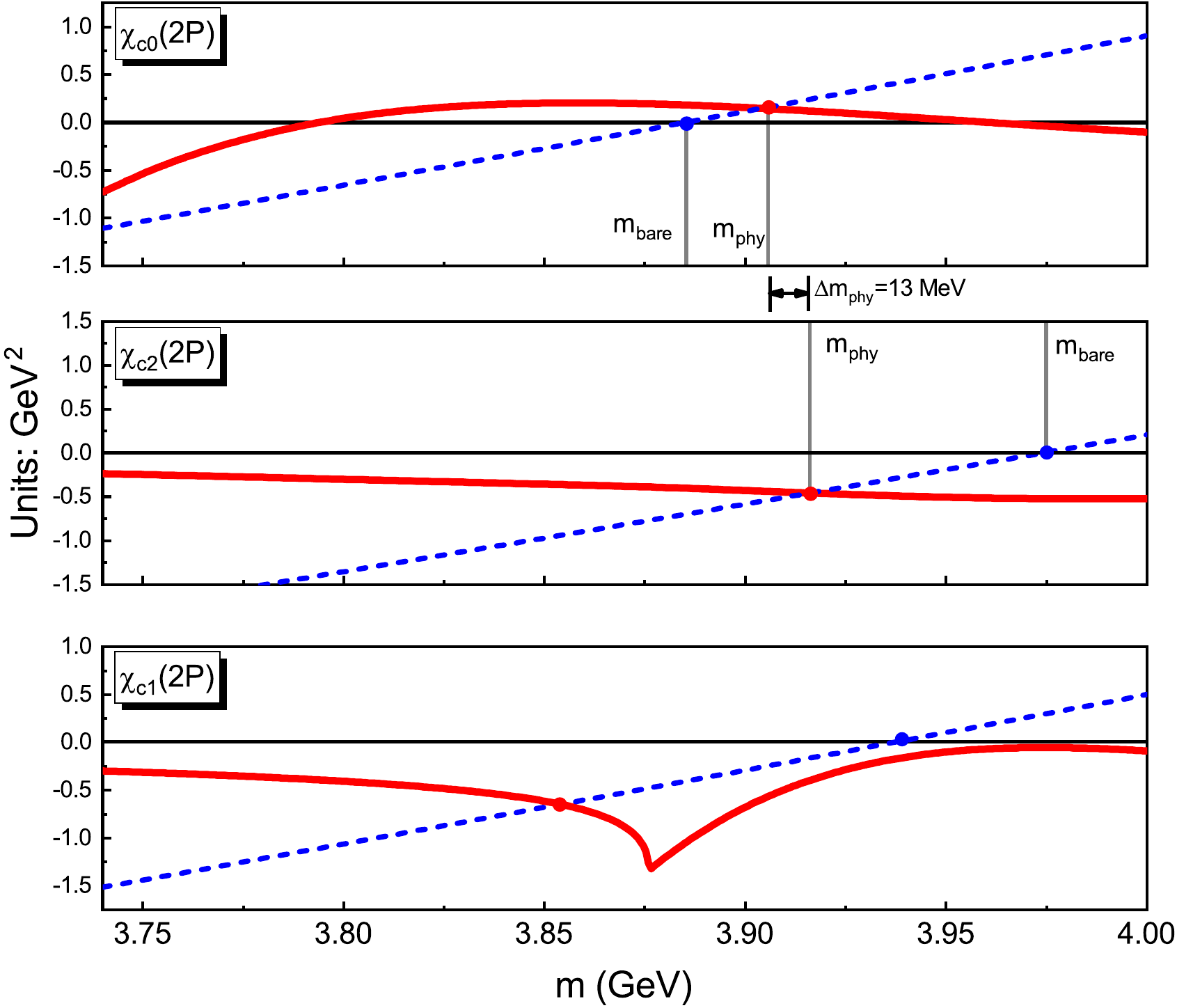}
\caption{The selfenergy function ${\rm Re}\Pi(m^2)$ of $\chi_{cJ}(2P)$ (red solid curve) and corresponding function $m^2-m_{\rm bare}^2$ dependent on $m$ (blue dot curve). The intersection of two curves is the solution of the equation $m_{\rm phy}^2=m_{\rm bare}^2+{\rm Re}\Pi(m_{\rm phy}^2)$, which corresponds to the physical mass.}
\label{2Pre}
\end{figure}

With the above preparation, we have no free parameter when presenting the result of the discussed $2P$ states of the charmonium family. As illustrated in Fig.~\ref{2Pre}, we may plot the dependence of the self energy function ${\rm Re}\Pi(m^2)$ and the corresponding function $m^2-m_{\rm bare}^2$ on $m$ for each discussed state. Then, we can find an intersection of these two curves, which corresponds to
an $m$ value. This $m$ value is the physical mass $m_{\rm phy}$ defined in Eq. (\ref{Narrow}).

Our result indicates:
\begin{itemize}

\item For $\chi_{c1}(2P)$, its physical mass is 3855 MeV, where the mass shift from the $D\bar{D}^*$ channel is -81 MeV, which shows that the unquenched effect is obvious. In this approach, the $1^{++}$ particle $X(3872)$ can be categorized as $\chi_{c1}(2P)$. Although there is small difference between the exact mass of $X(3872)$ and our result, we are still satisfied by our present result, since the result is obtained without free parameters and the low mass puzzle of $X(3872)$ is comprehensible.

\item For $\chi_{c0}(2P)$, the bare mass is 3885 MeV. After considering the unquenched effect, the mass shift is +19 MeV,  which is due to the $D\bar{D}$ channel contribution. Finally, the physical mass of $\chi_{c0}(2P)$ is 3904 MeV, which is consistent with the experimental width of $X(3915)$ observed in $\gamma\gamma\to \omega J/\psi$ \cite{Uehara:2009tx}. This can be seen later in the next subsections.

\item For $\chi_{c2}(2P)$, the unquenched effect from the $D\bar{D}$, $D\bar{D}^*$, and $D_s\bar{D}_s$ channels makes its physical mass lower down to 3917. Thus, assigning $Z(3930)$ existing in $\gamma\gamma\to  D\bar{D}$ \cite{Uehara:2005qd} as a $\chi_{c2}(2P)$ state is supported by our calculation of mass spectrum.

\end{itemize}
In Table \ref{Tabphy}, we summarize the above results for convenience of readers.

\begin{table}[htbp]
\footnotesize
\caption{The obtained physical masses for three $2P$ charmonium states. Additionally, their bare masses, widths  and $\delta m=m_{\rm phy}-m_{\rm bare}$ are given. {Here, these results are obtained by taking numerical spatial wave function listed in Fig. \ref{radial} and  Tables~\ref{C1}-\ref{C2} as input}.}
\label{Tabphy}
\renewcommand\arraystretch{1.1}
\begin{tabular*}{86mm}{@{\extracolsep{\fill}}ccccc}
\toprule[1pt]
\toprule[1pt]
State       &$m_{\rm bare}$ (MeV)&$m_{\rm phy}$ (MeV)&$\delta m$ (MeV)&$\Gamma$ (MeV)\\
\toprule[0.8pt]
$\chi_{c0}(2P)$&3885               &3904             & +19&23            \\
$\chi_{c1}(2P)$&3936               &3855              &-81&0             \\
$\chi_{c2}(2P)$&3974               &3917              &-57&26            \\
\bottomrule[1pt]
\bottomrule[1pt]
\end{tabular*}
\end{table}

We want to emphasize that  the mass gap between $\chi_{c2}(2P)$ and $\chi_{c0}(2P)$ can be decreased to only 13 MeV in our calculation, which shows that the small mass gap between $Z(3930)$ and $X(3915)$ (see Fig. \ref{tu2}) can be understood well.

Although this small mass gap between $Z(3930)$ and $X(3915)$ can be achieved in our unquenched model, we must face the serious problem. That is, before the present work, there are several theoretical calculations using the unquenched model \cite{Kalashnikova:2005ui, Li:2009ad, Ono:1983rd, Pennington:2007xr, Zhou:2013ada} as summarized in Sec. \ref{puzzle}. Why can we get this good result consistent with the experimental observation?

In the next subsection, we need to give an analysis to clarify this point, which makes our conclusion more convincing.

\subsection{How important is the node effect?}\label{dissect}

In this subsection, using Eqs. (\ref{Narrow}, \ref{ImPi}, \ref{MLS}, \ref{int}), we show how the node affects the decay width $\Gamma$ of $\chi_{c0}(2P)$. We also show the parameter $\beta$ dependence of masses and the mass gap between $\chi_{c0}(2P)$ and $\chi_{c2}(2P)$ so that the mass gap becomes smaller.

\begin{figure}[htbp]
\centering
\includegraphics[width=8.6cm,keepaspectratio]{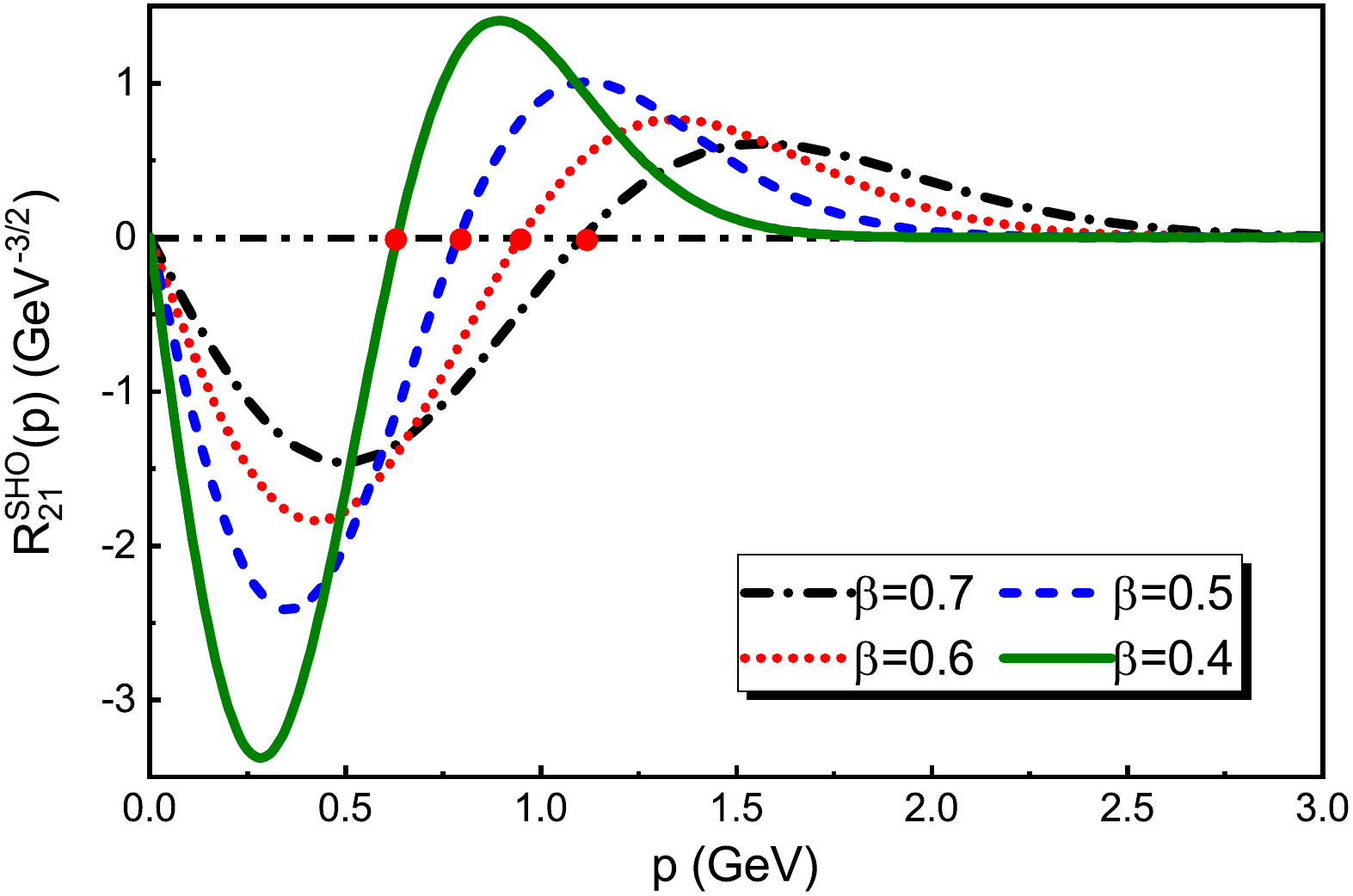}
\caption{The radial wave function of $\chi_{cJ}(2P)$ dependent on several typical values of $\beta$.
Here, the form of a radial wave function of $\chi_{cJ}(2P)$ is simply taken as the same as Eq. (\ref{wav0}).
The red points are the so-called node of a spatial wave function. $\beta$ is in unit of GeV.}
\label{WFbeta}
\end{figure}

For the $n$-$th$ radial excitation of a meson family, its spatial wave function $\psi_{nLM}(p)$ contains a radial one $R_{nL}(p)$ with $(n-1)$ nodes.
If taking a simple form like Eq. (\ref{wav0}) to express $R_{nL}(p)$, we can list its line shape dependent on $\beta$ as shown in Fig. \ref{WFbeta}, where we take $\chi_{cJ}(2P)$ state as an example.
For $\chi_{cJ}(2P)$ states, the principle quantum number is $n=2$, and the orbital angular momentum is $L=1$. At the node, a radial wave function can be separated into $R_{nL}(p)<0$ and $R_{nL}(p)>0$ parts. The position of a node changes with different $\beta$ values. 

\begin{figure}[htbp]
\centering
\includegraphics[width=8.6cm,keepaspectratio]{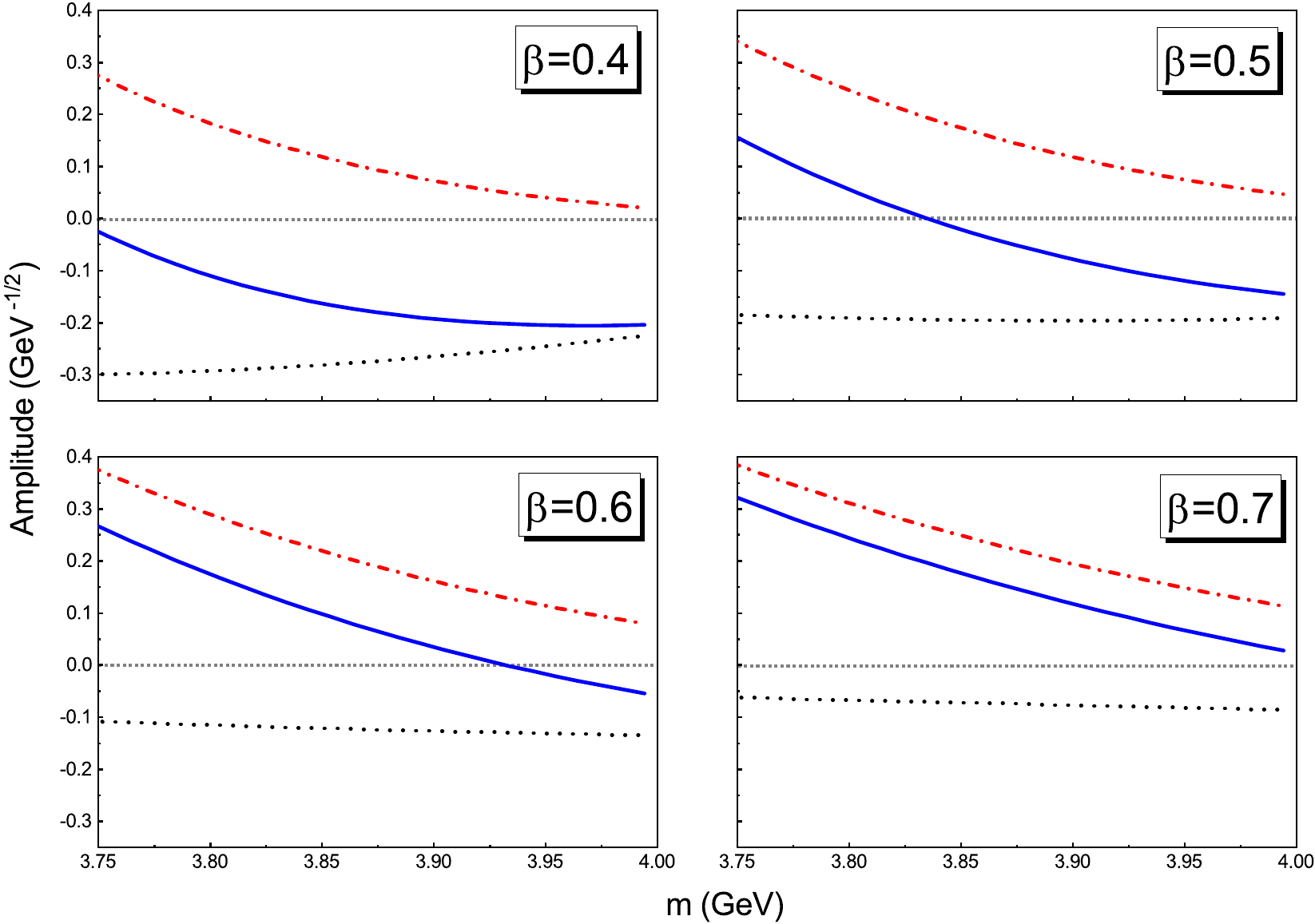}
\caption{The variation of $M^{LS}$ involved in $\chi_{c0}(2P)$ by changing the mass of $\chi_{c0}$ when taking $\beta=0.4$, $0.5$, $0.6$, $0.7$ GeV. Here, solid, dot, and dash-dot curves correspond to $M^{LS}$, $M^{LS}_{R_{nL}(p)>0}$, and $M^{LS}_{R_{nL}(p)<0}$, respectively.}
\label{node}
\end{figure}

Then, we apply this wave function to calculate the integral $I(P\hat{\bf z},m_1,m_2,m_3)$ given in  Eq.~(\ref{int}). Since it is the overlap of the finial and initial wave functions, the dependence of a node on $\beta$ directly results in the dependence of $I(P\hat{\bf z},m_1,m_2,m_3)$ on the $\beta$ value. To intuitively reflect this aspect, we take $\chi_{c0}(2P)$ affected by the $D\bar{D}$ channel as a typical example, where we still take a numerical wave function listed in Fig.~\ref{radial} for the final state $D$ meson as input. For $\chi_{c0}(2P)$, its radial wave function is defined by an SHO wave function given in Fig. \ref{WFbeta} to illustrate the $\beta$ dependence of $I(P\hat{\bf z},m_1,m_2,m_3)$. The integral in Eq.~(\ref{int}) is further rewritten as
\begin{equation}\label{inteRadial}
\small
\begin{aligned}
I(P\hat{\bf z}, m_{1}, m_{2}, m_{3})=&\int \mathrm{d}^{3} {\bf p}\; f({\bf p}, P\hat{\bf z})\psi_{n_{A} L_{A} M_{L_A}}({\bf p}),\\
=&\int_0^{\infty}\int_0^{4\pi}\left[f({\bf p}, P\hat{\bf z})Y_{L_{A}M_{L_A}}(\hat{\bf p})\right]R_{n_AL_A}(p) p^2 \mathrm{d}\Omega \mathrm{d}p,\\
=&\left(\int_0^{p_{\rm node}}R_{n_AL_A}(p) p^2\mathrm{d}p\
+\int_{p_{\rm node}}^{\infty}R_{n_AL_A}(p) p^2\mathrm{d}p\right)\\
&\times\int_0^{4\pi}\left[f({\bf p}, P\hat{\bf z})Y_{L_{A}M_{L_A}}(\hat{\bf p})\right] \mathrm{d}\Omega,
\end{aligned}
\end{equation}
where $f({\bf p}, P\hat{\bf z})$ represents the remaining parts other than $\psi_{n_{A} L_{A} M_{L_A}}({\bf p})$ in Eq.~(\ref{int}). $p_{\rm node}$ is the $p$ value corresponding to a node in a radial wave function
of $\chi_{c0}(2P)$.
The subscript $A$ in Eq.~(\ref{inteRadial}) is employed to label the $\chi_{c0}(2P)$ state.
In Eq.~(\ref{inteRadial}) , the integral $\int_0^{p_{\rm node}} R_{n_AL_A}(p) p^2\mathrm{d}p$ can {partially} cancel the contribution of $\int_{p_{\rm node}}^{\infty} R_{n_AL_A}(p) p^2\mathrm{d}p$. It is obvious that the node position becomes crucial to the result.
Then,
for Eq.~(\ref{MLS}), we may continue and define $M^{LS}=M^{LS}_{R_{nL}(p)<0}+M^{LS}_{R_{nL}(p)>0}$
according to Eq.~(\ref{inteRadial}), where $M^{LS}_{R_{nL}(p)<0}$ and $M^{LS}_{R_{nL}(p)>0}$ are related to $I(P\hat{\bf z},m_1,m_2,m_3)$ with $\int_0^{p_{\rm node}} R_{n_AL_A}(p) p^2\mathrm{d}p$ and $\int_{p_{\rm node}}^{\infty} R_{n_AL_A}(p) p^2\mathrm{d}p$, respectively.
In Fig.~\ref{node}, we present the dependence of $M^{LS}$ on the physical mass of $\chi_{c0}(2P)$ with four typical $\beta$ values, which will be applied to discuss the width of $\chi_{c0}(2P)$ state.
We find that the mass value corresponding to $M^{LS}=0$ changes with different $\beta$ values.

\begin{figure}[htbp]
\centering
\includegraphics[width=8.6cm,keepaspectratio]{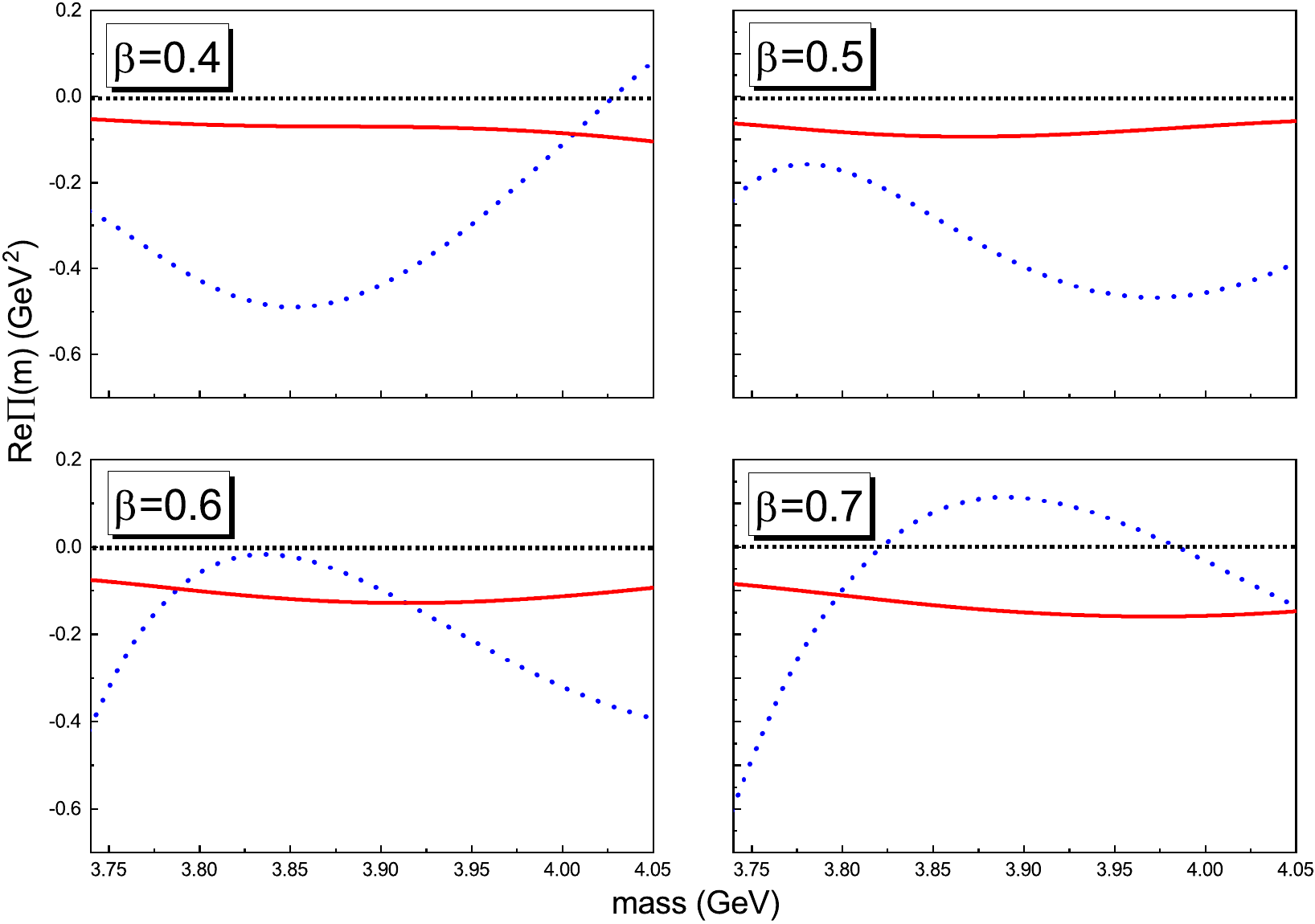}
\caption{Comparison of ${\rm Re}\Pi(m^2)$  of $\chi_{c2}(2P)$ (red solid curve) and $\chi_{c0}(2P)$ (blue dot curve) with four typical $\beta$ values.}
\label{re}
\end{figure}

The above analysis shows that the node effect should be emphasized. In Fig. \ref{re}, we further give
${\rm Re}\Pi(m^2)$ of $\chi_{c0}(2P)$ and $\chi_{c2}(2P)$ with different $\beta$ values, where the line shapes of ${\rm Re}\Pi(m^2)$ are dependent on a concrete $\beta$ value. Since ${\rm Re}\Pi(m^2)$ is a key step to determine the physical mass of $\chi_{c0}$ and $\chi_{c2}$, the physical mass of $\chi_{c0}$ and $\chi_{c2}$ must be dependent on the $\beta$ value (see
Table~\ref{result1} for more details).
\begin{table}[htbp]
\footnotesize
\caption{The unquenched results for $\chi_{cJ}(2P)$ with different $\beta$ values. $\beta$ is in unit of GeV.}
\label{result1}
\renewcommand\arraystretch{1.2}
\begin{tabular*}{86mm}{@{\extracolsep{\fill}}cccccc}
\toprule[1pt]
\toprule[1pt]
& &$\beta=$0.4&0.5&0.6&0.7\\
\toprule[0.8pt]
$\chi_{c0}(2P)$&$m_{\rm phy}$~(GeV)&3.824&3.849&3.877&3.900\\
$m_{\rm bare}=$3.885&$\Gamma$~(MeV)     &47&1&12&48\\
\toprule[0.8pt]
$\chi_{c1}(2P)$&$m_{\rm phy}$~(GeV)&3.879&3.871&3.859&3.849\\
$m_{\rm bare}=$3.937&$\Gamma$~(MeV)     &2&0&0&0\\
\toprule[0.8pt]
$\chi_{c2}(2P)$&$m_{\rm phy}$~(GeV)&3.932&3.922&3.912&3.906\\
$m_{\rm bare}=$3.974&$\Gamma$~(MeV)     &10&19&19&15\\
\bottomrule[1pt]
\bottomrule[1pt]
\end{tabular*}
\end{table}

We also find that the mass gap between $\chi_{c0}(2P)$
and $\chi_{c2}(2P)$ becomes smaller as the $\beta$ value increases.
In the former calculations by the unquenched models \cite{Pennington:2007xr, Zhou:2013ada}, the authors selected different wave functions as input, which results in the inconsistences among the obtained results.

In the present work, we take the GI model to get the numerical spatial wave function of the involved states. Before giving the inputs, we firstly
reproduce the mass spectrum of the well known charmonia.
This treatment avoids the uncertainty caused by spatial wave functions or the so-called $\beta$ value,  which also makes our conclusion to $\chi_{cJ}(2P)$ states reliable.
Finally, the reason why we may get small mass gap 
can be naturally explained by the above analysis. 

\subsection{The $\chi_{c0}(2P)$ state must be a narrow state!}

In Table \ref{Tabphy}, we also give our result of width of $\chi_{cJ}(2P)$ state.
For $\chi_{c2}(2P)$ state, the calculated width is 26 MeV, which is consistent with the experimental width of $Z(3930)$ ($\Gamma_{Z(3930)}=24\pm6$ MeV \cite{Tanabashi:2018oca}). This result supports the charmoniumlike state $Z(3930)$ to be a $\chi_{c2}(2P)$ state again.

In the following, we need to focus on the $\chi_{c0}(2P)$ state. Our unquenched calculation shows that
$\chi_{c0}(2P)$ should be a narrow state only with a width 23 MeV (see Table \ref{Tabphy}). If checking the resonance parameter of $X(3915)$, we find that our result overlaps with the measured width of $X(3915)$.
Here, the $\chi_{c0}(2P)$ state dominantly decays into a $D\bar{D}$ channel, which is a typical $S$-wave interaction. Since there is enough phase space for the $\chi_{c0}(2P)\to D\bar{D}$ decay, we usually guess that the partial decay width of $\chi_{c0}(2P)\to D\bar{D}$ is large before performing a realistic study. As indicated in Sec. \ref{dissect}, for the discussed $\chi_{cJ}(2P)$ states, the node effect is important.
When discussing the width of $\chi_{c0}(2P)$, the node effect on the width is obvious which can be reflected by the data from the third column in Table \ref{result1}.
Thus, assigning $X(3915)$ as a $\chi_{c0}(2P)$ state is fully possible. It is obvious that treating $X(3860)$ with a width $201$ MeV as $\chi_{c0}(2P)$ by Belle \cite{Chilikin:2017evr} cannot be supported by our present study. We also notice a theoretical work, where Wang, Liang and Oset indicated that it is questionable to assign $X(3860)$ as $\chi_{c0}(2P)$ \cite{Wang:2019evy} since the poor precise data of the Belle cannot rule out the existence of a $D\bar{D}$ bound/unbound state. 

We also noticed the recent LHCb's result of the $D\bar{D}$ invariant mass spectrum from the $pp$ collision \cite{Aaij:2019evc}. By analyzing the $D\bar{D}$ invariant mass spectrum, LHCb found a new narrow charmoniumlike state $X(3842)$ which can be a good candidate of $\psi(1^3D_3)$ state in the $J/\psi$ family. Accompanied by $X(3842)$, $\psi(3770)$ also exists in the measured $D\bar{D}$ invariant mass spectrum. Besides, there is a structure around 3.9 GeV. The LHCb Collaboration claim that this 3.9 GeV structure may correspond to $Z(3930)$ as $\chi_{c0}(2P)$ state. Thus, LHCb's data can be employed to search for charmonia with $D\bar{D}$ decay mode.

\begin{figure}[htbp]
\centering
\includegraphics[width=8.6cm,keepaspectratio]{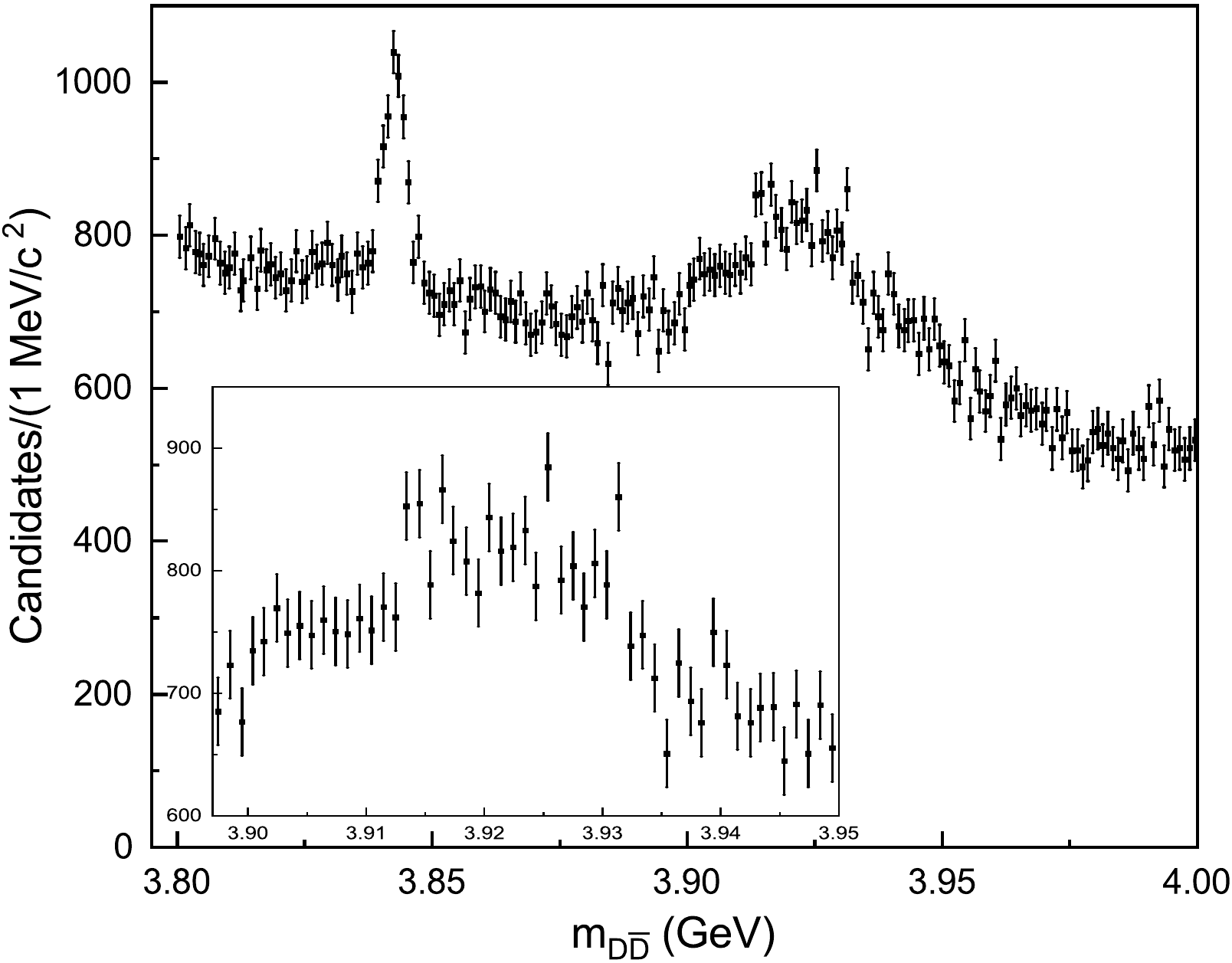}
\caption{The $D\bar{D}$ invariant mass spectrum from $pp$ collision in Ref.~\cite{Aaij:2019evc}}
\label{LHCbdata}
\end{figure}

In Fig. \ref{LHCbdata}, we collect the LHCb's data of the $D\bar{D}$ invariant mass spectrum, especially focusing on the 3.9 GeV structure.
We want to emphasize that this 3.9 GeV structure cannot be described by a simple Breit-Wigner
formula, and conjecture that this 3.9 GeV structure may contain at least two substructures according to our former analysis presented in Ref. \cite{Chen:2012wy}. In Ref. \cite{Chen:2012wy}, we once analyzed the structure around 3.9 GeV existing in the $D\bar{D}$ invariant mass spectrum from $\gamma\gamma\to D\bar{D}$ and indicated that this structure can be composed of $\chi_{c0}(2P)$ and $\chi_{c2}(2P)$.

We strongly suggest experimentalists to examine it. If our conjecture can be confirmed in experiment, one substructure may correspond to the $\chi_{c0}(2P)$ state and another denotes the $\chi_{c2}(2P)$ state. Observation of the $D\bar{D}$ decay mode of $X(3915)$ is the key point to finally
establish $X(3915)$ as $\chi_{c0}(2P)$ state.

We also want to comment on the Belle's result of $X(3860)$ \cite{Chilikin:2017evr} from $e^+e^-\to J/\psi D\bar{D}$ or the broad structure  $X(3840)$ with mass $3837.6\pm11.5$ MeV reported in Ref. \cite{Guo:2012tv} from $\gamma\gamma\to D\bar{D}$.
Since $X(3860)$ or $X(3840)$ exists in the $D\bar{D}$ structure, there should exist their explicit signal in the LHCb's data of the $D\bar{D}$ invariant mass spectrum. Unfortunately, we cannot find any evidence either of $X(3860)$ or $X(3840)$
in the $D\bar{D}$ invariant mass spectrum released by LHCb \cite{Aaij:2019evc}. This fact cannot be
evaded by the authors in Refs. \cite{Guo:2012tv} if treating $X(3860)$ \cite{Chilikin:2017evr} or the so-called $X(3840)$ as $\chi_{c0}(2P)$. Here, it is time to seriously check whether the broad structures $X(3860)$ \cite{Chilikin:2017evr} and $X(3840)$ \cite{Guo:2012tv} are due to resonance contribution or background, which will be a crucial task left to experimentalists.

Finally, we should state our opinion on the $\chi_{c0}(2P)$ state: {\it $\chi_{c0}(2P)$ must be a narrow state and the charmoniumlike state $X(3915)$ is a good candidate of $\chi_{c0}(2P)$ without any doubt}.

\section{Summary}\label{sec5}

Since the observation of $J/\psi$ in 1974, the charmonium family has become abundant. In the past 17 years, the charmoniumlike $XYZ$ states have been reported, which not only provides a good chance to explore exotic hadronic states but also gives us an opportunity to identify a missing charmonium.
However, the road to identify a missing charmonium is not smooth. A typical example is $X(3915)$ discovered in $\gamma\gamma\to \omega J/\psi$ by Belle \cite{Uehara:2009tx}. In the former work, the Lanzhou group indicated that $X(3915)$ is a good candidate for the $\chi_{c0}(2P)$ state \cite{Liu:2009fe}. Later, BaBar confirmed that the $J^{PC}$ quantum number is $0^{++}$ by performing angular momentum analysis \cite{Lees:2012xs}. According to this result, the 2013 version of PDG \cite{Beringer:1900zz} labeled $X(3915)$ as $\chi_{c0}(2P)$. However,
some theoretical groups proposed three problems against such an assignment (see the review in Sec. \ref{introduction}). Among these problems, it has been a crucial task we have to face how to explain the small mass gap between $X(3915)$ and $Z(3930)$.

In this work, we have seriously studied the possibility of $X(3915)$ as $\chi_{c0}(2P)$. For the discussed $\chi_{cJ}(2P)$ states, they are above the $D\bar{D}$ and $D\bar{D}^*$ thresholds. Thus, a coupled-channel effect should be considered when performing such a study, which is a typical unquenched picture for hadrons. Based on an unquenched quark model, we have calculated the mass spectrum of three $\chi_{cJ}(2P)$ states. To avoid the uncertainty from input parameters, we have fixed the $\gamma$ value and have taken numerical spatial wave functions of the involved states calculated by the GI model. Having carried out the GI model calculation, we have reproduced the masses of the well-established charmonia.
Having done the above treatment, no free parameter has existed in our calculation.  Our results have shown that the mass difference between $\chi_{c0}(2P)$ and $\chi_{c2}(2P)$ is 13 MeV, which is very close to the mass gap between $X(3915)$ and $Z(3930)$. Of course, the masses of $X(3915)$ and $Z(3930)$ have been reproduced in the present work.
For letting the reader to convince our result, we have given an analysis to explain why we can reach such good results different form the former unquenched model calculation, where the importance of node effects due to spatial wave functions of $2P$ charmonium is explicitly indicated.

Besides mass spectrum analysis to support the assignment of $X(3915)$ as $\chi_{c0}(2P)$, we have also calculated the width of $\chi_{c0}(2P)$ to be 23 MeV. Such a value is also consistent with the experimental data of $X(3915)$, which further enforces the possibility of $X(3915)$ as $\chi_{c0}(2P)$. Especially, in this work we have emphasized that $\chi_{c0}(2P)$ should be a narrow state.

To finally establish $X(3915)$ as $\chi_{c0}(2P)$, the search for $X(3915)\to D\bar{D}$ is crucial. In Ref.~\cite{Chen:2012wy}, the Lanzhou group proposed that the 3.9 GeV structure corresponding to $Z(3930)$ in the $D\bar{D}$ invariant mass spectrum of $\gamma\gamma\to D\bar{D}$ should be composed of two substructures, which gives a solution of the dominant $D\bar{D}$
channel of $X(3915)$ missing in experiments. Recent LHCb's data of the $D\bar{D}$ invariant mass spectrum from $pp$ collision \cite{Aaij:2019evc} can again support the above proposal since the 3.9 GeV structure existing in LHCb's data cannot be depicted by one structure. We strongly encourage an experimental study of the detailed structure around 3.9 GeV found by LHCb from the  $D\bar{D}$ invariant mass spectrum data.

{
Before making a final conclusion $X(3915)$ as $\chi_{c0}(2P)$, we still need to face the so-called consistency problem existing in two estimated branching ratios of $\mathcal{B}(\chi_{c0}(2P)\to \omega J/\psi)$, which was proposed in Ref. \cite{Olsen:2019lcx}. Here, Olsen adopted two approaches to estimate $\mathcal{B}(\chi_{c0}(2P)\to \omega J/\psi)$: (1) assuming that both $X(3915)$ from the $\gamma\gamma\to J/\psi\omega$ process and $Y(3940)$ from $B^+\to J/\psi\omega K^+$ \cite{Abe:2004zs} are originated from the same state $\chi_{c0}(2P)$, one expects $\mathcal{B}(B^+\to K^+Y(3940))=\mathcal{B}(B^+\to K^+\chi_{c0}(2P))\leq\mathcal{B}(B^+\to K^+ \chi_{c0}(1P))$. Then, one obtains the lower limit $\mathcal{B}(Y(3940)\to J/\psi\omega)=\mathcal{B}(\chi_{c0}(2P)\to J/\psi\omega)>0.14$, where the experimental values $\mathcal{B}(B^+\to K^+\chi_{c0}(1P))=1.5^{+0.15}_{-0.14}\times 10^{-4}$ \cite{Agashe:2014kda}
and $\mathcal{B}(B^+\to K^+ Y(3930))\times\mathcal{B}(Y(3940)\to J/\psi \omega)=3.0^{+0.6+0.5}_{-0.5-0.3}\times 10^{-5}$ \cite{delAmoSanchez:2010jr,Aubert:2007vj} were employed in this estimate; (2) applying the relation from the quenched potential model \cite{Olsen:2019lcx}
\begin{eqnarray}
\frac{\Gamma(\chi_{c0}(2P)\to \gamma\gamma)}{\Gamma(\chi_{c2}(2P)\to \gamma\gamma)}=\frac{\Gamma(\chi_{c0}(1P)\to \gamma\gamma)}{\Gamma(\chi_{c2}(1P)\to \gamma\gamma)}=4.4\pm0.6,\label{uu}
\end{eqnarray}
one gets an upper limit $\mathcal{B}(\chi_{c0}(2P)\to J/\psi \omega)<8.1\%$ with the experimental value $\Gamma(X(3915)\to \gamma\gamma)\times\mathcal{B}(X(3915)\to\omega J/\psi)=54\pm 9$ eV \cite{Agashe:2014kda} as input. In this work, taking this opportunity, we want to give comments on the above estimate of the branching ratio of $\chi_{c0}(2P)\to J/\psi\omega$:
\begin{itemize}
\item Although there exists similarity of the resonance parameters of $X(3915)$ and $Y(3940)$, this treatment of $X(3915)$ as the same as $Y(3940)$ is not acceptable in the whole community (see a review article \cite{Chen:2016qju,Liu:2013waa}). In fact, $Y(3940)$ from $B^+\to J/\psi\omega K^+$ \cite{Abe:2004zs} is a good candidate of a $D^*\bar{D}^*$ molecular state as indicated in Ref. \cite{Liu:2009ei}. 
Thus, this value of $\mathcal{B}(B^+\to K^+ Y(3940))\times\mathcal{B}(Y(3940)\to J/\psi \omega)$ cannot be applied to estimate the branching ratio of $\chi_{c0}(2P)\to J/\psi\omega$.

\item Equation (\ref{uu}) is only valid under the framework of a quenched quark model. For these higher charmonia with mass above the threshold of a charmed meson pair, the hadronic loop contribution should be considered in calculating their decays. In Ref. \cite{Chen:2013yxa} , the Lanzhou group performed a realistic study of $X(3915)\to J/\psi\omega$ and $Z(3930)\to J/\psi \omega$, which occurs via intermediate hadronic loops composed of charmed mesons. The result shows that the partial decay width of $\chi_{c2}(2P)\to J/\psi \omega$ is at least one order of magnitudes smaller than that of $\chi_{c0}(2P)\to J/\psi\omega$ \cite{Chen:2013yxa}. It is obvious that the relation shown in Eq. (\ref{uu}) is violated by a hadronic loop effect when discussing higher charmonia $\chi_{c0}(2P)$ and $\chi_{c2}(2P)$. Thus, the estimate of the upper limit of a branching ratio of $\chi_{c0}(2P)\to J/\psi\omega$ in Ref. \cite{Olsen:2019lcx}  is questionable. 

\end{itemize}
As illustrated above, we would like to emphasize that the consistency problem raised in Ref. \cite{Olsen:2019lcx} does not exist. 
Of course, investigating the $\chi_{c0}(2P)\to J/\psi\omega$ decay in the near future will still be an interesting issue. }

We hope that the present work can provide valuable information to clarify the messy situation of identifying the candidate of $\chi_{c0}(2P)$. In the following years, experimentalists should dedicate themselves to this tough problem accompanied by theorists, where LHCb and Belle II will still play the main force role.

\section*{Acknowledgement}
This project is partly supported by the China National Funds for Distinguished Young Scientists under Grant No. 11825503 and the National Program for Support of Top-notch Young Professionals.


\end{document}